\documentclass[aip,jcp,amsmath,amssymb,preprint]{revtex4-1}
\usepackage{dcolumn}
\usepackage{bm}
\usepackage{graphicx} 
\usepackage{epstopdf} 
\usepackage{color}

\begin{document}  
\preprint{AIP/123-QED} 
\title[Brownian motion in time-dependent logarithmic potential]
{Brownian motion in time-dependent logarithmic potential: \\ Exact results for dynamics and
first-passage properties} 
\author{Artem Ryabov} 
\email{rjabov.a@gmail.com} 
\author{Ekaterina Berestneva} 
\author{Viktor Holubec} 
\affiliation{ 
Charles University in Prague, 
Faculty of Mathematics and Physics, 
Department of Macromolecular Physics,
V Hole{\v s}ovi{\v c}k{\' a}ch 2, 180~00 Praha~8, Czech Republic}
\date{\today} 
\begin{abstract} 
The paper addresses Brownian motion in the logarithmic potential with time-dependent strength, $U(x,t) = g(t) \log(x)$, subject to the absorbing boundary at the origin of coordinates. Such model can represent kinetics of diffusion-controlled  reactions of charged molecules or escape of Brownian particles over a time-dependent entropic barrier at the end of a biological pore. We present a simple asymptotic theory which yields the long-time behavior of both the survival probability (first-passage properties) and the moments of the particle position (dynamics). The asymptotic survival probability, i.e., the probability that the particle will not hit the origin before a given time, is a functional of the potential strength. As such it exhibits a rather varied behavior for different functions $g(t)$. The latter can be grouped into three classes according to the regime of the asymptotic decay of the survival probability. We distinguish  1.~the regular (power-law decay), 2.~the marginal (power law times a slow function of time), and 3.~the regime of enhanced absorption (decay faster than the power law, e.g., exponential). Results of the asymptotic theory show good agreement with numerical simulations. 
\end{abstract} 
\maketitle  
\section{Introduction}  

Logarithmic potential $U(x)=g\log(x)$, frequently appears as an effective potential in diffusion models of biophysics and chemical physics. For instance, it is used to model interactions among polyelectrolytic polymers,\cite{Manning1969} it arises as an entropic term in free energy costs for unzipping of DNA macromolecules,\cite{PS1, PS2, FogedbyMetzler2007, KaiserNovotny2014} as an entropic potential stemming from interactions of colloids with walls of narrow channels,\cite{Zwanzig1991, Reguera2001, Reguera2012} and as a ``potential'' for momentum diffusion in optical lattices.\cite{LutzRenzoniNATPHYS2013, Kessler2010, LutzBarkaiPRL2011,  KesslerBarkaiPRL2012, KesslerBarkaiPRE2012, LeibovichBarkai2015} 

Other prominent examples include long-range interacting particles,\cite{Bouchet2005}  self-gravitating Brownian particles,\cite{Chavanis2002, Chavanis2007} diffusion of eigenvalues of random matrices known as Dyson's Brownian motion,\cite{Dyson1962, Spohn1987} and interaction of tracers in driven lattice gasses.\cite{SchutzEPL2005}  
As for single-particle problems, diffusion in logarithmic potential has recently regained significant attention due to its intriguing slow relaxation\cite{KesslerBarkaiPRL2012, KesslerBarkaiPRE2012, LeibovichBarkai2015, SchutzPRE2011, SchutzJSTAT2011} and rather nontrivial first-passage properties.\cite{Bray2000, Fogedby2003, Martin2011} Moreover, combining the logarithmic and the parabolic potentials\cite{Giorno, Giampaoli, Lo, Fogedby2003} one obtains exactly solvable models for stochastic resonance\cite{WioResonance} and for energetics of a Brownian particle in an asymmetric optical trap.\cite{RyabovLog2013}

In probability theory, the Brownian motion in logarithmic potential with a constant strength $g$ is known as the  \emph{Bessel process},\cite{Yor2003} which can be defined as the distance to the origin of a $(g+1)$-dimensional Brownian motion.\cite{KarlinTaylor1, KarlinTaylor2} In economy, scaling transformation of the Bessel process generate the so called Constant Elasticity of Variance model,\cite{LinetskyCEV2010, Makarov2010} see Ref.~\onlinecite{Martin2011} for more references in this direction. 

In the present work we consider the overdamped Brownian motion in the logarithmic  potential with \emph{time-dependent} strength
\begin{equation}
\label{LogPot}
U(x,t)=g(t)\log(x).
\end{equation} 
Then the position $\mathbf{X}(t)$ of the Brownian particle evolves according to the Langevin equation 
\begin{equation}
\label{Langevin}
{\rm d} \mathbf{X}(t) = - \frac{g(t)}{\mathbf{X}(t)} {\rm d}t + \sqrt{2D}\, {\rm d}\mathbf{W}(t),
\end{equation}
where $D$ is the diffusion constant and ${\bf W}(t)$ denotes the standard Wiener process. Our primary interest is to characterize the \emph{first-passage time} ${\bf T}$, which is the time when the particle hits the coordinate $x=0$ for the first time.\cite{BrayMajumdarScher, Redner}
 The statistical properties of ${\bf T}$ are  described by the \emph{survival probability} $S(t)$, also called persistence \cite{BrayMajumdarScher}, which gives the probability that the particle has not hit the origin by the time $t$, $S(t)={\rm Prob}\!\left\{ {\bf T}>t \right\}$. 

First-passage problems in chemical physics are used to model rates of diffusion-limited reactions,\cite{Redner, Havlin, HavlinAdvances} exit kinetics of Brownian particles diffusing out from biological micropores,\cite{RC2012} or rates of any other process activated by diffusion. 
In our case one can imagine that Eq.~(\ref{Langevin}) provides a rough model for the exit of a diffusing particle from a micropore with a time-dependent entropic barrier at the pore end (see Refs.~\onlinecite{Zwanzig1991, Reguera2001, Reguera2012}, where it has been shown that the entropic potential is proportional to the logarithm of the pore width). 
At the same time, Eq.~(\ref{Langevin}) can describe diffusion of a small mobile ion in the presence of a polyelectrolyte (long charged polymeric molecule).\cite{Manning1969} In the latter case, ${\bf X}(t)$ is understood as a distance of the ion from the charged polymer chain and the logarithmic potential can model the Coulomb potential energy of the small mobile ion [see Eqs.~(4) and (15) in Ref.~\onlinecite{Manning1969}]. The strength $g(t)$ of the logarithmic potential is then proportional to the density of charges on the polyelectrolyte. One can assume that this charge density depends on time due to the interaction of the polyelectrolyte with other small mobile ions. For instance the ions can ``condense'' on the polyelectrolyte reducing its charge density.\cite{Manning1969}

To the best of our knowledge, there is no exact solution for the proposed model (\ref{Langevin}) when the potential strength $g(t)$ depends arbitrarily on time. The present paper partially fills this gap. We present exact  \emph{asymptotic} description of both the first-passage properties and the particle dynamics. 
Our main result is the exact asymptotic expression (\ref{Sasy}) for the survival probability $S(t)$. 
This expression suggests that functions $g(t)$ can be divided into three classes according to the resulting asymptotic decay of $S(t)$ (see three dynamic regimes in Sec.~\ref{sec:threeregimes}).
The result  (\ref{Sasy})  also determines the long-time asymptotic behavior of the probability density function (PDF) for the particle position, which yields asymptotic expressions for any moment $\left<{\bf X}^{n}(t) \right>$ as discussed in Sec.~\ref{sec:moments}. 

Finally it should be noted that, from a general perspective, the proposed model belongs to a class of first-passage problems with time-dependent conditions. Exact results in this field, even long-time asymptotic ones, are rather rare. Analytical progress is feasible for a free (no potential) Brownian motion in domains of varying size,\cite{Breiman1967, Novikov1981, KrapivskyRedner1996, MartinLof1998, GodrecheLuck1999, GodrecheLuck2001, Newman2001, BraySmith2007F, BraySmith2007, Forsberg2008, SimpsonPLOSONE2015, Simpson2015, SimpsonJCP2015} for the Bessel process with drifting position of the singularity,\cite{DeLong1981, Ronzhin1986, Enriquez2008, Salminen2011} for a Brownian motion in time-dependent piecewise linear potentials\cite{Agudov2001} (see also Ref.~\onlinecite{Dubkov2004}  for a related concept of ``nonlinear relaxation time''), when the linear potential depends exponentially on time \cite{UrdapilletaPRE2011, Urdapilleta2011, Urdapilleta2012} and for a weak potential\cite{Linder2004} treated as perturbation. Our present findings extend this list by a nontrivial example with a rather rich physical behavior. 

We start the presentation by the precise definition of the model in terms of a Fokker-Planck equation supplemented by appropriate boundary conditions (Sec.~\ref{sec:model}). Essential properties of the Bessel process (constant $g$) are recapitulated in Sec.~\ref{sec:Bessel}. Sec.~\ref{sec:theory} comprises our main results: exact asymptotic solution of the Fokker-Planck equation (\ref{FokkerPlanck}) (Sec.~\ref{sec:exact}) and classification of functions $g(t)$ into three groups (three dynamic regimes in Sec.~\ref{sec:threeregimes}). Asymptotic evolution of the particle position is discussed in Sec.~\ref{sec:moments}. Appendix describes the simulation algorithm used to check predictions of the asymptotic theory.

\section{\label{sec:model}Definition of the Model}  
Consider diffusion of a particle in a semi-infinite one-dimensional interval $(0,\infty)$ with an \emph{absorbing boundary} at the origin. Initially, at $t=0$, the particle is situated at $x_0$, $x_0 >0$. For $t>0$ the particle diffuses in the logarithmic potential (\ref{LogPot}), hence its position evolves according to the Langevin equation (\ref{Langevin}). The boundary at the origin is perfectly absorbing, i.e., if the particle hits the origin, it is immediately absorbed with probability one. Thus the hitting time of the origin equals the time ${\bf T}$ of the first passage to the origin. 

 The Fokker-Planck equation [corresponding to Eq.~(\ref{Langevin})] for the PDF of the particle position reads
\begin{equation} 
\label{FokkerPlanck}
\frac{\partial }{\partial t}p(x,t) = 
D \frac{\partial^{2} }{\partial x^{2}} 
p(x,t) 
+ \frac{\partial }{\partial x} 
\left[ \frac{g(t)}{x} 
p(x,t) \right]. 
\end{equation} 
The absorbing boundary at the origin is represented by the absorbing boundary condition $p(0,t) = 0$ and the initial condition is $p(x,0)=\delta(x-x_{0})$.  
The survival probability $S(t)$, that is the probability that ${\bf T} >t$, follows from $p(x,t)$ after the spatial integration   
\begin{equation}
S(t) = \int_{0}^{\infty}p(x,t) {\rm d} x,
\end{equation}
since $S(t)$ equals the probability that, at time $t$, the particle can be found somewhere in the interval $(0,\infty)$.

\section{\label{sec:Bessel}Static phase diagram ($\boldsymbol g \boldsymbol = \textrm{const.}$)}  
Let us now recapitulate properties of the Bessel process (constant $g$), which will serve as a starting point in the following discussion. Derivations of discussed facts can be found e.g.\ in Ref.~\onlinecite{Bray2000}.

\begin{figure}[t!] 
\begin{center} 
\includegraphics[width=0.5\columnwidth]{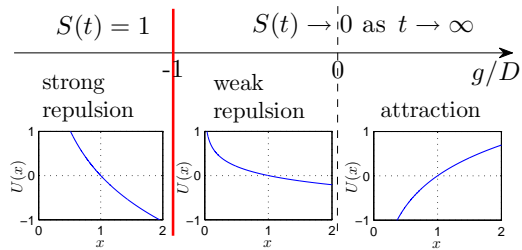} 
\caption{Static phase diagram (constant $g$). In the region $g/D \leq -1$  the logarithmic potential $U(x)=g\log(x)$  is strongly repulsive as compared to the intensity of thermal fluctuations, therefore the particle never hits the origin. In all other cases ($g/D >-1$), the survival probability decays to zero as $t\to \infty$. Its asymptotic decay is given in Eq.\ (\ref{BesselAsy}). 
Insets illustrate the potential for the individual regions of the phase diagram, $U(x)$ is plotted for $g=-1.5$, $g=-0.3$ and $g=1$ (assuming that $D=1$).} 
\label{fig:PhaseDiag}
\end{center}
\end{figure}

Dynamics of the model with a constant $g$ exhibits two qualitatively different regimes. The master parameter which determines qualitative features of the dynamics is the ratio $g/D$. When $g/D\leq-1$, the potential is strongly repulsive as compared to the intensity of thermal fluctuations. Due to the strong repulsion, the Brownian particle \emph{never} hits the origin $x=0$ and its survival probability equals to one for any time.   
On the other hand, when $g/D>-1$, i.e., for the weakly repulsive potential $[g/D \in (-1,0)]$, for the free Brownian motion $(g=0)$, and for the attractive potential ($g > 0$), the particle eventually hits the origin with probability one. Hence for $g/D>-1$ we have $S(t)\to 0$ as $t\to\infty $. 
In Fig.~\ref{fig:PhaseDiag}, these dynamic regimes are summarized in what we call the static (constant $g$) phase diagram of the model. 
We can translate these conclusions for the ion diffusion in the presence of a polyelectrolytic molecule (see Introduction). When $g>0$, the small mobile ion and the polyelectrolyte possess opposite charges, thus it is not surprising that the ion will eventually be attracted to the polymeric chain. However, even when the ion and the polyelectrolyte bear charges of the same sign ($g<0$), the ion can hit the polyelectrolyte provided  the temperature is large enough and/or the charge density on the polyelectrolyte (which is proportional to $g$) is small, i.e., when $0>g/D>-1$.

In order to trace back the origin of such a nontrivial behavior notice that (for a constant $g$) the Langevin equation (\ref{Langevin}) will preserve its form after the ``Brownian'' rescaling of coordinate and time, ${\bf X}(t)\to\alpha^{1/2}{\bf X}(t)$, $t\to\alpha t$. 
\footnote{The simplest way how to see the scaling behavior of the drift is to set $D=0$ and take $g<0$. Then the particle position will grow as the square root of time ${\bf X}(t)\sim t^{1/2}$, i.e., with the dynamic exponent that controls also the diffusive spreading of the particle PDF (when $D>0$).} 
Since the two competing tendencies -- the diffusive spreading of the particle PDF and the drift of the particle away from the origin -- scales with time in the very same way, it is the ratio of their amplitudes, $g/D$, which decides the type of resulting dynamics.

In the present paper we will not discuss the case of the strong repulsion where $g/D<-1$. We will focus on $g$ such that $g/D \geq -1$, which includes both the region $g/D>-1$, where the survival probability decays to zero and the boundary case $g=-D$, where the survival probability equals to one for all times.  
The exact PDF of the particle position for $g/D \geq -1$ reads\cite{Bray2000}
\begin{equation} 
\label{BesselPDF}
p_{\rm B}(x,t) = \frac{x_0}{2Dt}\left( \frac{x}{x_0 } \right)^{\frac{1}{2}\left(1-g/D\right)}
\!\!\!\!\! \exp\!\left(- \frac{x^{2}+x_{0}^{2}}{4Dt} \right)
{\rm I}_{\nu}\!\left( \frac{x x_0 }{2Dt} \right) ,
\end{equation}
where the order of the modified Bessel function of the first kind $\nu$ depends on the potential strength, $\nu= \frac{1}{2} (1+g/D)$. The subscript ``B'' will mark all quantities related to the Bessel process.  

The survival probability $S_{\rm B}(t)$ is obtained from the PDF $p_{\rm B}(x,t)$ after the spatial integration,  
$S_{\rm B}(t) = \int_{0}^{\infty}p_{\rm B}(x,t) {\rm d}x$. For $g/D>-1$ the integral may be expressed in terms of the ``upper'' incomplete gamma function, $\Gamma(a,z) = \int_{z}^{\infty} t^{a-1}{\rm e}^{-t} {\rm d}t$, as follows 
\begin{equation}
\label{BesselS}
S_{\rm B}(t) = 1- 
\frac{\Gamma\!\left[ \frac{1}{2} (1+g/D) , \frac{x_{0}^{2} }{4Dt} \right]}
{\Gamma\!\left[\frac{1}{2}(1+g/D)  \right]}.
\end{equation}
For $4Dt \gg x_{0}^{2} $, i.e., when the diffusion length is much larger than the initial distance of the particle from the origin, Eq.~(\ref{BesselS}) simplifies to the power law 
\begin{equation}
\label{BesselAsy}
S_{\rm B}(t) \sim 
\frac{ \left( \left. x_{0}^{2} \right/ 4D \right)^{\frac{1}{2}\left(  1+  {g / D} \right)}}{\Gamma\left[\frac{1}{2}(3+g/D) \right]} \,
 t^{- \frac{1}{2}\left( 1 + g/D  \right)} , \qquad t\to \infty.
\end{equation} 
Notice that the time-independent prefactor of the long-time asymptotic form (\ref{BesselAsy}) depends on the initial position~$x_{0}$. This long-time memory effect illustrates that the first-passage time ${\bf T}$, which is characterized by $S_{\rm B}(t)$, is a \emph{functional} of the stochastic process ${\bf X}(t)$ (see reviews \onlinecite{BrayMajumdarScher, MajumdarFunctional} for more details). Thus even in the long-time limit the asymptotic survival probability should bear information about the whole history of the process ${\bf X}(t)$. The latter fact is clearly visible in our asymptotic result (\ref{Sasy}), where the asymptotic survival probability depends on the integral of $g(t)$. 

We remind that the PDF (\ref{BesselPDF}) is valid also for the boundary case which corresponds to $g=-D$. In this case, however, the PDF (\ref{BesselPDF}) is normalized to one for all times, i.e., $S_{\rm B}(t)=1$ for all $t$. 

\section{\label{sec:theory}Dynamic phase diagram [$\boldsymbol g \boldsymbol  = \boldsymbol g(\boldsymbol t)$]}  

If the strength of the logarithmic potential  depends on time, $g=g(t)$, then, to the best of our knowledge, the exact solution of the Fokker-Planck equation (\ref{FokkerPlanck}) is unknown. 
However,  using the results for the Bessel process, it is possible to find the \emph{asymptotically exact} solution of this equation for arbitrary $g(t)$. In this section we present the theory and then discuss the asymptotic behavior of the survival probability $S(t)$. The asymptotic expression for $S(t)$ suggests that functions $g(t)$ can be divided into three classes (three dynamic regimes in Sec.~\ref{sec:threeregimes}) according to the resulting asymptotic decay of $S(t)$. 

\subsection{Asymptotically exact solution of the Fokker-Planck equation \label{sec:exact}}  

The central quantity for the following asymptotic theory is the PDF of the particle position given that the time of absorption ${\bf T}$ is greater than $t$, $p\left(x,t| \mathbf{T}>t \right)$. This PDF  \emph{conditioned on nonabsorption before} $t$, is defined as 
\begin{equation}
\label{CondPDFdef}
p\left(x,t| \mathbf{T}>t \right) = \frac{p(x,t)}{S(t)}. 
\end{equation}
The  approximation made below is motivated by a remarkable property of the conditional PDF of the Bessel process: its asymptotic behavior for \emph{any} constant $g$ is given by the Rayleigh distribution
\begin{equation} 
\frac{p_{\rm B}(x,t)}{S_{\rm B}(t)} \sim
\frac{x}{2 Dt} \, {\rm e}^{-\left. x^{2}\right/4Dt},
\qquad t\to \infty,
\label{BesselCond}
\end{equation}
which can be verified dividing the power series representation of PDF (\ref{BesselPDF}) by that of the survival probability~(\ref{BesselS}). 
Thus regardless the actual strength of the logarithmic potential $g$, the conditional PDF is exactly the same as that for the free Brownian motion.\cite{RyabovChvosta2014} This can be understood on physical grounds. When $t$ is large, the PDF conditioned on non-absorption describes statistical properties of \emph{long-living} trajectories. Such trajectories are typically situated far from the origin. At the same time, at large distances from the origin the force ($-g/x$) acting on the particle is vanishingly small and hence the long-time limit (\ref{BesselCond}) does not dependent on the amplitude~$g$.  

As a starting point of the approximative scheme we extrapolate the above observation to the case of time-dependent potential strength $g(t)$. That is, in the long-time limit we approximate the conditional PDF $p\left(x,t| \mathbf{T}>t \right)$ by the Rayleigh distribution (\ref{BesselCond}) and hence for $p(x,t)$ we assume the asymptotic approximation \begin{equation}
\label{Ansatz} 
p(x,t) \sim 
S(t) \frac{x}{2 Dt} \, {\rm e}^{-\left. x^{2}\right/4Dt},
\quad t \to \infty,
\end{equation} 
where $S(t)$ is an unknown function yet to be determined. 
In order to derive $S(t)$, which gives us the leading asymptotic behavior of the survival probability, we require that the Ansatz (\ref{Ansatz}) for $p(x,t)$ satisfies the exact Fokker-Planck equation (\ref{FokkerPlanck}) subject to the absorbing boundary condition $p(0,t)=0$.

If we insert the Ansatz (\ref{Ansatz}) into the Fokker-Planck equation (\ref{FokkerPlanck}) it turns out that all $x$-dependent terms cancel. The final result is that the Ansatz (\ref{Ansatz}) satisfies the exact Fokker-Planck equation (subject to the absorbing boundary condition) if and only if the asymptotic survival probability $S(t)$ satisfies the first order ordinary differential equation 
\begin{equation}
\frac{\rm d}{\rm d t}S(t) = 
- \frac{D+ g(t) }{2 D t} S(t). 
\label{ODES} 
\end{equation} 
Solution of Eq.~(\ref{ODES}) yields asymptotic behavior of the survival probability $S(t)$, 
\begin{equation}
S(t) \sim A \exp\!\left( - \int^{t}_{t_0} 
\frac{D+ g(t') }{2 D t' } 
 {\rm d}t' \right),
 \quad t\to\infty,
 \label{Sasy}
\end{equation}
where the lower limit $t_0$, $t_{0}<t$, is arbitrary since the constant ($t$-independent) term which originates from this limit can always be included into the $t$-independent prefactor $A$ (the actual value of the prefactor cannot be determined from the above theory). In the whole paper we can simply set $t_{0}=0$ except in Fig.~\ref{fig:logrelax}, where $t_0=2$ in order to avoid possible divergence caused by $g(t)$, see the caption.

Finally let us stress that the above derivation holds also for $g(t)$ which lies on the phase boundary $g(t)=-D$, cf. Fig.~\ref{fig:PhaseDiag}.

\subsection{\label{sec:threeregimes}Three dynamic regimes}  

The asymptotic survival probability (\ref{Sasy}) is the functional of the potential strength. As such it can exhibit a rather varied behavior for different functions $g(t)$. Actually, Eq.~(\ref{Sasy}) suggests that different $g(t)$ can be grouped into three classes according to the type (regime) of the asymptotic decay of $S(t)$. Specifically, we distinguish  1.~\emph{the regular regime}, 2.~\emph{the marginal regime}, and 3.~the regime of \emph{enhanced absorption}.

\emph{1.\ The regular regime} is the only one where the asymptotic survival probability (\ref{Sasy}) can \emph{decay with time as the power law} with a well defined exponent. 
This regime occurs when two necessary conditions are satisfied. (i) The first condition is that the potential strength tends to the limit $g_{\infty}$, $-D\leq g_{\infty} <\infty$, as $t\to\infty$. 
That is, we can express $g(t)$ as 
\begin{equation}
\label{gregular}
g(t) = g_{\infty} + \varepsilon(t), 
\end{equation}
where the residue $\varepsilon(t)$ vanishes when $t\to \infty$. Then for $S(t)$ we obtain
\begin{equation}
\label{Seps}
S(t) \sim A t^{-\frac{1}{2}(1+g_{\infty}/D)} 
\exp\!\left( - \int^{t}_{t_0} 
\frac{\varepsilon(t') }{2 D t' } 
 {\rm d}t' \right),
 \quad t\to\infty. 
\end{equation} 
(ii) The second condition is that the exponential in the above equation should converge to a constant as $t\to\infty$. 
To this end, the integral in the exponential must be bounded from above, $\int^{t}_{t_0 }\left[ \varepsilon(t')/t' \right]{\rm d}t' < \infty$, as $t\to\infty$.
For example this is true for $\varepsilon(t)$ which decays to zero as an arbitrary power of time, $\varepsilon(t)\sim t^{-a}$, $a>0$. Then the integral in the exponential tends to a constant as $t^{-a}/a$ with increasing time and the exponential in Eq.~(\ref{Seps}) does not influence the leading asymptotic decay of $S(t)$. Other prominent example is when $\varepsilon(t)$ oscillates around zero with a given period, say $\varepsilon(t)=B\sin(\omega t)$, for which $\int_{t_{0}}^{t} \left[\sin(\omega t')/t'\right] {\rm d}t'$ again converges to a constant.

\begin{figure}[t!]
\begin{center}
\includegraphics[width=0.5\columnwidth]{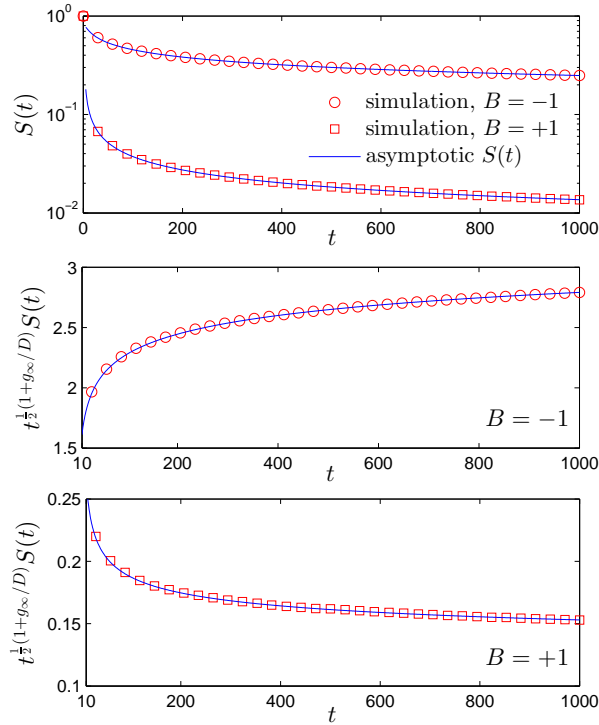} 
\caption{Decay of the survival probability in the marginal regime for $g(t)=-0.3+B/\log(t+5)$, $t\in [0,10^{3}]$, and for two values of $B$, $B=-1$ (the prefactor in the asymptotic survival probability (\ref{Smarginal}) is $A\approx 1.062$) and $B=+1$ ($A\approx 0.4021$). The survival probability appears to be much larger for $B=-1$, as predicted by the logarithmic term in Eq.~(\ref{Smarginal}), which is shown in lower two panels. Other parameters used: $x_{0}=1$, $D=1$; simulated data were averaged over $10^{6}$ trajectories.} 
\label{fig:marginal} 
\end{center} 
\end{figure}

In the regular regime, when $g_{\infty}/D > -1$, the survival probability decays as the power law 
$t^{-\frac{1}{2}(1+g_{\infty}/D)}$, 
that is, exactly as that for the Bessel process (\ref{BesselAsy}) with $g=g_{\infty}$. On the other hand, even though the exponents in the both asymptotic formulas can be equal, the time-independent prefactors are in general different. The reason for this difference is that the constant $A$ in Eq.\ (\ref{Seps}) depends on the whole history of the process including the initial condition and the transient behavior of $g(t)$. We are not able to determine the constant $A$ within our asymptotic theory. In examples below, when we compare the analytical result (\ref{Sasy}) with simulated data, $A$ is treated as the (only) fitting parameter.

Besides the power law relaxation when $g_\infty /D>-1$, the second possibility can occur within the regular regime. Namely, if $g(t)$ tends to the phase boundary ($g_\infty=-D$, cf.\ the static phase diagram in Fig.~\ref{fig:PhaseDiag}), the survival probability (\ref{Seps}) converges to a \emph{nonzero} value equal to $A$ as $t\to\infty$.

\begin{figure}[t!]
\begin{center}
\includegraphics[width=0.5\columnwidth]{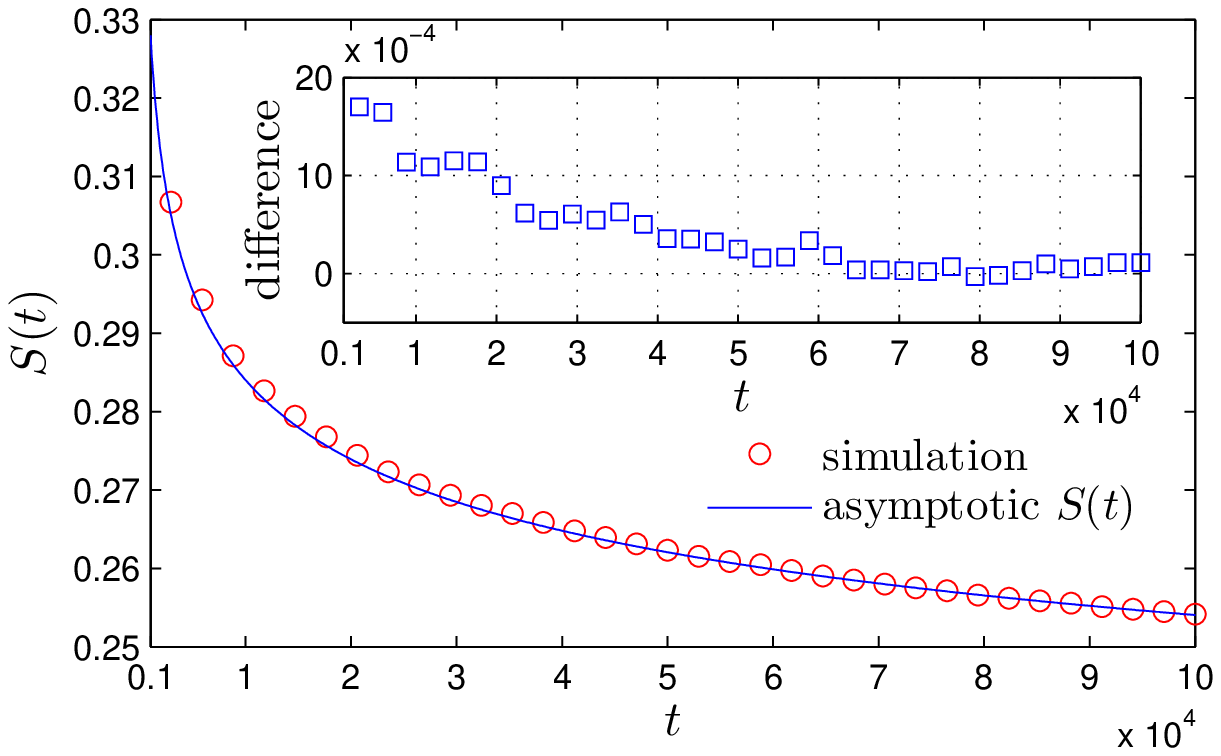} 
\caption{Slow logarithmic decay of $S(t)$ according to Eq.~(\ref{Smarginal}) with $g_{\infty}=-D$, $B=1$ and $D=1$. In simulations (symbols) we have used $g(t)=-1+1/\log(t)$, $t\in [2 ,10^{5}]$, $x_0=1$; $10^{5}$ trajectories were generated in total.  The inset shows the difference between the simulated $S(t)$ and the exact asymptotic behavior (\ref{Smarginal}). Apparently the higher-order corrections to Eq.~(\ref{Smarginal}) also decay logarithmically with time when $g_\infty =-D$.} 
\label{fig:logrelax} 
\end{center}
\end{figure} 

\emph{2.\ In the marginal regime} the assumption (i) used to define the regular regime still holds and $g(t)$ is again given by Eq.~(\ref{gregular}), whereas, the assumption (ii) concerning the fast decay of $\varepsilon(t)$ is no longer valid. Consequently, the exponential in Eq.~(\ref{Seps}) no longer tends to a constant as $t\to\infty$. Instead, it behaves as a \emph{slow function of time}, which modifies the asymptotic decay of $S(t)$.
A particularly simple example of the function $\varepsilon(t)$ which leads to the marginal behavior reads
\begin{equation}
\label{epslog}
\varepsilon(t) \sim \frac{B}{\log(t)},\quad t\to\infty.
\end{equation}
In this case, when $g_{\infty}/D>-1$, the asymptotic survival probability tends to zero as the power law multiplied by the power of the logarithm:
\begin{equation}
\label{Smarginal}
S(t) \sim A t^{-\frac{1}{2}(1+g_{\infty}/D)} 
\left[
\log(t)
\right]^{-B/2D}, 
 \quad t\to\infty. 
\end{equation}
The logarithmic multiplicative term is clearly visible in simulated data, see two lower panels in Fig.~\ref{fig:marginal}. The sign of $B$ significantly influences the absolute value of $S(t)$ at finite times. When $B>0$, $g(t)$ is always higher than its $g_\infty$ which leads to more rapid decrease of $S(t)$. For $B<0$, the value of $g(t)$ at any finite time is closer to the phase boundary as compared to $g_\infty$. This results in much slower decrease of $S(t)$ as demonstrated in the uppermost panel of Fig.~\ref{fig:marginal}.

A particularly interesting situation arises when $g(t)$ asymptotically approaches the phase boundary $g_{\infty}=-D$. Then the power-law prefactor in Eq.~(\ref{Seps}) is absent and the asymptotic behavior of $S(t)$ is determined solely by the decay of $\varepsilon(t)$.
The resulting super slow decay of the survival probability for the logarithmic $\varepsilon(t)$ from Eq.~(\ref{epslog}) is demonstrated in Fig.~\ref{fig:logrelax}. Eventually $S(t)$ approaches zero, which is in sharp contrast to what we observe in the regular regime, where $S(t)$ tends to a nonzero value equal to $A$, $A<1$.

\begin{figure}[t!] 
\begin{center} 
\includegraphics[width=0.5\columnwidth]{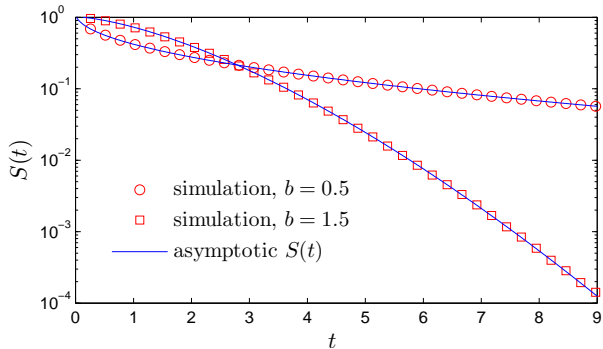} 
\caption{Simulated $S(t)$ (symbols) for $g(t)=-D+t^{b}$, compared to the asymptotic result (\ref{SgincrExp}) (solid lines) for two different values of the exponent $b$,  $b=0.5$ (the fitted prefactor is $A\approx 1.141$) and $b=1.5$ ($A\approx 1.015$). Remaining parameters read $D=1$, $x_{0}=0.1$. Simulated data were averaged over $10^{6}$ trajectories.} 
\label{fig:exponent}
\end{center} 
\end{figure} 

\emph{3.\ The regime of enhanced absorption} occurs when $g(t)\to\infty$ as $t\to \infty$. 
In this case the potential becomes strongly attractive for large times (we move to the right in the static phase diagram in Fig.~\ref{fig:PhaseDiag} as $t\to\infty$). Its strength can be decomposed as  
\begin{equation} 
g(t) = g_{0} + \eta(t), 
\end{equation}
where $g_{0}$, $g_{0}\geq -D$, denotes the value of $g(t)$ at the initial time and $\eta(t)\to\infty$ when $t\to\infty$. For the asymptotic survival probability (\ref{Sasy}) we obtain 
\begin{equation}
S(t) \sim A t^{-\frac{1}{2}(1+g_{0}/D)} \exp\!\left( - \int^{t}_{t_0 } 
\frac{\eta(t')}{2 D t' } 
 {\rm d}t' \right),
 \quad t\to\infty.
 \label{Sgincr} 
\end{equation}

In both the marginal and the regular regime $S(t)$ decreases at most as the power of time [possibly multiplied by a slow function, see Eq.~(\ref{Smarginal})] with the exponent being controlled by the limit $g_{\infty}$. 
Presently, since $\eta(t)$ grows with time, the overall asymptotic behavior is determined by the exponential in Eq.~(\ref{Sgincr}). The pre-exponential power law, which is now controlled by the initial value $g_{0}$, plays a rather minor role.  

\begin{figure}[t!] 
\begin{center} 
\includegraphics[width=0.5\columnwidth]{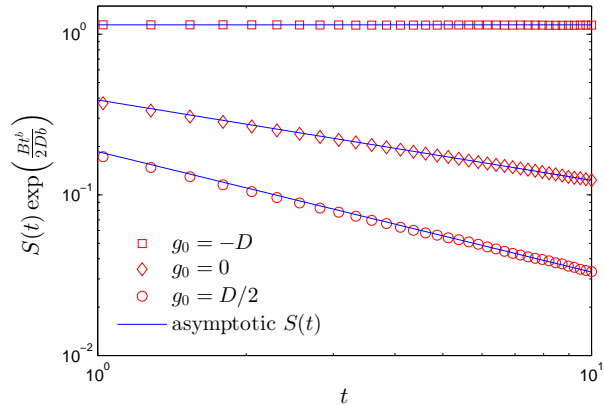} 
\caption{Log-log plot of the pre-exponential factor in the asymptotic result (\ref{SgincrExp}) for three values of $g_0$. Simulation of $S(t)$ (symbols) was performed for $g(t)=g_0+Bt^{b}$ with $B=1$ and $b=1/2$. Other parameters used: $D=1$, $x_{0}=0.5$ when $g_0 =0$ and $g_0=D/2$, and $x_0 = 0.1$ when $g=-D$. The fitted prefactor $A$ reads 
$A\approx 1.144$ for $g_0 = -D$, 
$A\approx 0.390$ for $g_0 = 0$ and  
$A\approx 0.186$ for $g_0 = D/2$. 
Simulated data were averaged over $10^{6}$ trajectories.} 
\label{fig:preexponential} 
\end{center} 
\end{figure}

To illustrate the enhanced decay of $S(t)$, let us assume a power-law increase of the potential strength
\begin{equation}
g(t) = g_{0} + B t^{b}, \quad B>0, \quad b>0,
\end{equation}
which yields the exponential asymptotic behavior  
\begin{equation}
S(t) \sim A t^{-\frac{1}{2}(1+g_{0}/D)} \exp\!\left( -  
\frac{B t^{b}}{2D b } \right),
 \quad t\to\infty.
 \label{SgincrExp}
\end{equation}
Value of the exponent $b$ is decisive in the long-time limit, as demonstrated in Fig.~\ref{fig:exponent}, where the analytical result (\ref{SgincrExp}) is compared with simulated data for two values of $b$. On the other hand, the initial value $g_{0}$ modifies just the  pre-exponential factor. Its effect is however still well visible, see Fig.~\ref{fig:preexponential} which shows the results for different values of $g_{0}$. 


\subsection{Failure of naive adiabatic approximation \label{sec:adiabatic}}  

Assume that the potential strength $g(t)$ varies rather slowly as compared to the diffusive spreading of the particle PDF. For such slow functions $g(t)$, one can guess a straightforward approximation for $S(t)$.  Namely, one can assume that at each time $t$ the particle effectively moves in a constant potential. 
This is the basic idea of the adiabatic approximation, which is frequently a rather successful method how to obtain results for time-dependent problems.\cite{Talkner2004, Talkner2005} In the spirit of the adiabatic approximation we can try to substitute the slow function $g(t)$ for the constant $g$ in the exact PDF for the Bessel process (\ref{BesselPDF}).  The survival probability is then given by the spatial integral of such PDF. We thus obtain  
\begin{equation} 
S_{\rm ad}(t) = 1- 
\frac{\Gamma\!\left[ \frac{1}{2} (1+g(t)/D) , \frac{x_{0}^{2} }{4Dt} \right]}
{\Gamma\!\left[\frac{1}{2}(1+g(t)/D) \right]},
\end{equation} 
which is the analogue of the survival probability (\ref{BesselS}) but now with time-dependent $g(t)$.
However, even for very slow functions $g(t)$, this naive approximation gives different results as compared to the exact formula (\ref{Sasy}). Again the reason for this difference is that $S(t)$ should reflect the complete history of the process, thus it is a functional of $g(t)$ and not just a function of its instantaneous value.

\section{\label{sec:moments}Moments of the particle PDF}  

The asymptotic theory presented in Sec.~\ref{sec:exact} gives us also the leading asymptotic behavior of all moments of the PDF $p(x,t)$. 
From the definition of the conditional PDF (\ref{CondPDFdef}) it follows that the $n$th moment of  $p(x,t)$ can be expressed as the product  
\begin{equation}
\label{momentsdef}
\left<\mathbf{X}^{n}(t) \right> = S(t) \left<\mathbf{X}^{n}(t) \right>_{\mathbf{T}>t} ,
\end{equation}
of the survival probability $S(t)$ and the $n$th \emph{conditional moment} $\left<\mathbf{X}^{n}(t) \right>_{\mathbf{T}>t}$. The latter is computed from PDF conditioned on nonabsorption before $t$,
\begin{equation}
\left<\mathbf{X}^{n}(t) \right>_{\mathbf{T}>t} = 
\int_{0}^{\infty} x^{n}p\left(x,t|\mathbf{T}>t \right) {\rm d}x .
\end{equation}
In the long-time limit the conditional moments are given by those of the Rayleigh distribution, cf.\ the right-hand side of Eq.~(\ref{BesselCond}). Thus we have
\begin{equation}
\label{momentsn}
\left<\mathbf{X}^{n}(t) \right>_{\mathbf{T}>t} \sim \Gamma(1+n/2) (4Dt)^{n/2},
\qquad t\to \infty.
\end{equation} 
For instance for the first two moments of the particle position we obtain the asymptotic expressions 
\begin{eqnarray}
&& \left< \mathbf{X}(t) \right> \sim S(t) \sqrt{\pi D t} , \\
&& \left< \mathbf{X}^{2}(t) \right> \sim S(t) 4D t , 
\end{eqnarray}
where, for a specific $g(t)$, the asymptotic survival probability $S(t)$ should be computed from Eq.\ (\ref{Sasy}) as discussed in the preceding section~\ref{sec:theory}. 
Alternatively, one can obtain main asymptotic behavior of any moment $\langle {\bf X}^{n}(t)  \rangle $ from the moment-generating function $\chi(\xi,t)$, $\chi(\xi,t)=\langle {\rm e}^{\xi {\bf X}(t) } \rangle$. Evaluation of the mean $\langle {\rm e}^{\xi {\bf X}(t) } \rangle$ with respect to the PDF (\ref{Ansatz}) yields 
\begin{equation} 
\chi(\xi,t) \sim S(t) 
\left[1 + \xi \sqrt{\pi Dt } \,{\rm e}^{Dt \xi^{2}} 
{\rm erfc}\! \left(- \xi \sqrt{Dt } \right) \right],
\end{equation}
which, after expansion into a power series in $\xi$, gives  the moments (\ref{momentsn}) multiplied by the survival probability $S(t)$, cf.\ Eq.~(\ref{momentsdef}).

\section{\label{sec:truncated}Regularized logarithmic potential}  

Let us now consider a first passage time to the point $a$ which is not exactly at the origin, $0<a<x_0$. In order to reach $a$ the Brownian particle has to overcome a \emph{finite} (time-dependent) potential barrier, in contrast to the previous case $a=0$, where the potential barrier is infinite. Such ``regularization'' of the logarithmic potential introduces an additional length scale which could significantly affect behavior of the survival probability. However, it turns out that the main asymptotic behavior of $S(t)$ (now the probability that the particle has not reached the boundary at $x=a$ before $t$), as given by Eq.~(\ref{Sasy}), remains valid also for the present regularized case. In the asymptotic formula (\ref{Sasy}) the regularization affects just the value of the constant prefactor $A$.

\begin{figure}[t!] 
\begin{center} 
\includegraphics[width=0.5\columnwidth]{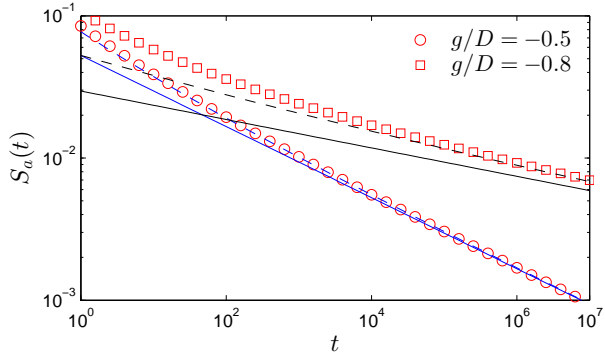} 
\caption{
Log-log plot of $S_{a}(t)$ for two values of $g$. The survival probability $S_{a}(t)$ has been obtained by the numerical inversion of its Laplace transform (\ref{Slaplace}) (symbols). Solid lines show only the first term of the asymptotic representation (\ref{Saasy}),  dashed lines show both terms from (\ref{Saasy}). Other parameters used: $D=1$, $a=0.5$ $x_{0}=0.6$.} 
\label{fig:constantg} 
\end{center} 
\end{figure} 

New physical effects, which stem from the regularization of the potential, are not reflected in the time-dependence of the principal asymptotic part of $S(t)$. Instead, {\em an anomalous  relaxation towards this asymptotic behavior} is observed when $a>0$ (as compared to {\em the normal} relaxation for $a=0$). The relaxation is described by higher-order terms of the asymptotic expansion of $S(t)$, which can be studied analytically for a constant $g$.
In this case, the survival probability, say $S_{a}(t)$, can be derived from the transition PDF~(\ref{BesselPDF}) by a standard approach described e.g.\ in Refs.~\onlinecite{Siegert,SiegertDarling}. Its Laplace transform,  
\begin{equation}
\label{Slaplace} 
\tilde{S}_{a}(s) = 
\frac{1}{s}\left[1-
\left(\frac{x_{0}}{a}\right)^{\nu} 
\frac{ {\rm K}_{\nu}(x_{0}\sqrt{s/D}) }{{\rm K}_{\nu}(a\sqrt{s/D})}
 \right], 
\end{equation} 
is valid for any $g/D$, i.e., for the whole phase diagram in Fig.~\ref{fig:PhaseDiag}. 
The order $\nu$ of the modified Bessel function of the second kind is determined by the ratio of the potential strength and the diffusion coefficient, $\nu=\frac{1}{2}(1+g/D)$. 
In the region of the phase diagram which is close to the phase boundary, i.e., when  $g/D \in (-1,1)$, the first two terms of the long-time asymptotic series read\cite{DoetschBOOK} \footnote{The long-time asymptotic behavior of $S_{a}(t)$ is linked to the asymptotic representation of $\tilde{S}_{a}(s)$ near the singularity closest to the abscissa of convergence. The singularity is the branch point at $s=0$. Thus expanding $\tilde{S}_{a}(s)$ into the power series near $s=0$ gives us the asymptotic behavior (\ref{Saasy}).\cite{DoetschBOOK}}
\begin{equation} 
\label{Saasy} 
S_{a}(t) \sim 
\frac{x_{0}^{2 \nu}-a^{2 \nu}}{\Gamma(1+\nu)}\left( \frac{1}{4Dt}\right)^{\nu}+
a^{2\nu} \frac{x_{0}^{2 \nu}-a^{2 \nu}}{\Gamma(1-2\nu)} 
\left(\frac{\Gamma(1-\nu)}{\Gamma(1+\nu)}\right)^{2} 
\left( \frac{1}{4Dt}\right)^{2 \nu}.
\end{equation} 
Contrary to this, when the potential is strongly attractive ($g/D >1 $) the second term of the asymptotic representation of $S_{a}(t)$ would decay with time as $1/t^{\nu +1}$.  

First of all notice that indeed the leading asymptotic term in (\ref{Saasy}) is, up to the constant prefactor, identical to that given in Eq.~(\ref{BesselAsy}) for the Bessel process. 
On the basis of the leading asymptotic term only, one could conclude that the regularized problem ($a>0$) is rather similar to the case when $a=0$. 
However, the second term in (\ref{Saasy}) reveals that for a finite time $t$ this conclusion is false. When $g/D$ is close to $-1$ this second term vanishes extremely slowly with time. 
Thus for a successful quantitative comparison of $S_{a}(t)$ (or of the simulated data) with the asymptotic representation (\ref{Saasy}) it is essential to include also this term (and possibly higher order terms). Fig.~\ref{fig:constantg} demonstrates this fact for two values of $g/D$. Note that for $g/D=-0.8$, the second term is still visible at times $t\approx 10^{7}$. 

The {\em anomalous relaxation} towards the main asymptotic behavior is absent when $a=0$. In the limit $a\to 0$ all slow terms vanish and the asymptotic series of $S_{a}(t)$ reduces to that for the Bessel process: $S_{\rm B}(t)\sim c_0 t^{-\nu}\left( 1+c_{1}/t + c_{2}/t^{2} + ...\right)$, where the first-order correction vanishes ($1/t$)-times faster than the leading asymptotic behavior. 

\begin{figure}[t!] 
\begin{center} 
\includegraphics[width=0.5\columnwidth]{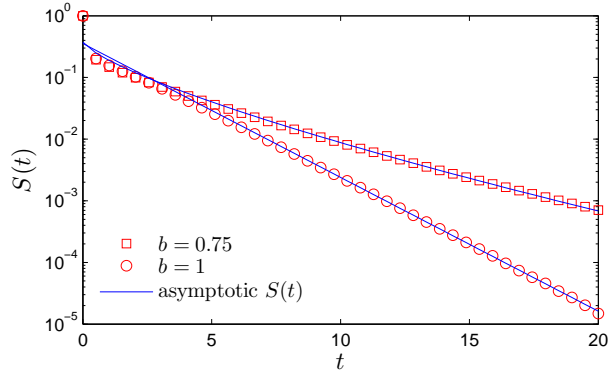} 
\caption{The simulated survival probability for the regularized logarithmic potential (symbols) with strength $g(t)=-D+t^{b}$ and for two values of $b$. Other parameters used: $a=0.3$, $x_0 =0.4$, $D=1$. Simulated data were averaged over $10^{7}$ trajectories; asymptotic $S(t)$ is given by (\ref{Sasy}).}  
\label{fig:Saexp} 
\end{center} 
\end{figure} 

Anomalous relaxation of the same kind is present when the potential strength depends on time. We have numerically simulated the process for several $g(t)$ in each dynamical regime. The simulations support the above observation. Namely, when $g(t)/D$ is close to the phase boundary, we observed a slow relaxation similar to that in Fig.~\ref{fig:constantg}. When $g(t)/D$ is far from the phase boundary [e.g. $g(t)/D>1$], then the asymptotic result (\ref{Sasy}) shows satisfactory agreement with simulated data. Typical example of the latter situation is demonstrated in Fig.~\ref{fig:Saexp} (the regime of enhanced absorption), where $g(t)$ departs from the phase boundary and grows as a power of time. Notice that for small times the simulated data differ significantly from the asymptotic curve. This is in sharp contrast to what we have observed for $a=0$, cf.\ Fig.~\ref{fig:exponent}, where the asymptotic survival probability (\ref{Sasy}) agrees with simulation even for rather small times.

\section{\label{sec:concluding}Concluding remarks and open questions}
Can a Brownian particle overcome the time-dependent logarithmic potential barrier (\ref{LogPot}) and hit the origin? If so, what are the statistical properties of the hitting (first-passage) time?  When the potential strength is \emph{constant}, the problem is exactly solvable and the origin can be reached only if $g/D>-1$ as summarized in the static phase diagram in Fig.~\ref{fig:PhaseDiag}. In this case the survival probability decays as the power law (\ref{BesselAsy}). For a \emph{time-dependent} potential strength the problem becomes more complex and, moreover, we have no exact solution of the Fokker-Planck equation (\ref{FokkerPlanck}) at our disposal. 
In the present paper we have presented simple long-time asymptotic theory which leads to the long-time  solution (\ref{Ansatz}) of this equation.  
The asymptotic PDF  (\ref{Ansatz}) factorizes into two terms. One term is the $x$-dependent Rayleigh distribution which is the conditional PDF for ensemble of long-living trajectories. Such trajectories are typically situated far from the origin, hence the conditional PDF contains no trace of the initial position $x_{0}$ and of the logarithmic singularity at $x=0$. 
The other term, the asymptotic survival probability $S(t)$ given by Eq.~(\ref{Sasy}), bares information about both the initial position $x_0$ and the behavior of the function $g(t')$ for all times $t'$ before $t$. In Eq.~(\ref{Sasy}) the initial condition is contained in the time-independent prefactor $A$ [similarly as in Eq.~(\ref{BesselAsy}) for a constant $g$]. The result (\ref{Sasy}) predicts rather rich asymptotic behavior of $S(t)$ depending on concrete form of $g(t)$. The three regimes of relaxation of $S(t)$ can be identified, see Sec.~\ref{sec:threeregimes}. Having derived Eq.~(\ref{Sasy}) one can return to the Ansatz (\ref{Ansatz}) and thus obtain the  asymptotic dynamics of the particle position, see Sec.~\ref{sec:moments}. 
Let us now turn to open questions which naturally stem from the presented results. 

First of all a few further remarks on the nature of approximation used are in order. 
Although the Ansatz for the long-time PDF (\ref{Ansatz}) was motivated by the time-independent case, it turned out to be surprisingly successful when predicting the time-dependent behavior (see comparison with numerics for various $g(t)$ in Figs.~2-5). 
This success stems from the fact that the differential equation (\ref{ODES}) \emph{is closed}, i.e., it contains no $x$-dependent terms, and thus the Ansatz (\ref{Ansatz}) itself satisfies the exact Fokker-Planck equation without any further approximation.  
On the other hand, on physical grounds one can expect that the key approximation of the conditional PDF (\ref{CondPDFdef}) by the Rayleigh distribution should at long times hold for a rather broad class of potentials. For instance for potential barriers localized near origin and for decreasing potentials $U(x,t)=g(t)/x^{\alpha}$, $\alpha>0$. The both examples yield force terms in the Langevin equation which decay faster than ($-g/x$) with increasing $x$. However, in these more general cases the resulting differential equation for the asymptotic $S(t)$ is no longer closed and hence further approximations are needed. In future work it could be very interesting to extend the present theory to these potentials and/or to justify the present approximation rigorously, e.g., as an expansion in a small parameter. 
Such extensions of the present method could result in a unifying treatment of several rather nontrivial first-passage problems in time-dependent potentials so common in biophysics and chemical physics.

\begin{acknowledgments}
The authors are grateful to J{\' a}n {\v S}omv{\' a}rsky for discussions related to origin-crossing of the Bessel process. 
\end{acknowledgments}

\appendix
\section{Simulation} 
\label{AppendixS} 

In order to verify asymptotic results we need the algorithm which fulfills two requirements: (i) it treats exactly the neighborhood of the origin where the potential diverges; (ii) the average time step in the algorithm is large enough so that we could efficiently simulate the long-time regime. In order to meet both (i) and (ii), we have divided the code into two parts. Firstly, when the particle is far from the origin (when $x\geq 1$ in our code, this condition is satisfactory for all sets of parameters used in Figs.~\ref{fig:marginal}-\ref{fig:preexponential}) we integrate the Langevin equation (\ref{Langevin}) by simple Euler-Maruyama method.\cite{KloedenBOOK} Secondly, near origin (for $x<1$ in our code) we have generated a new position from the exact PDF (\ref{BesselPDF}) using the inverse transformation method.\cite{RossBOOK} In the latter part the time-step $\Delta t$ must be such that $g(t)$ can be treated as constant in $(t,t+\Delta t)$. For more details related to simulation of Bessel-like processes see Refs.~\onlinecite{Makarov2010, Martin2011}.

\bibliography{paper}
\end{document}